\documentclass[twocolumn,prb,aps,floats,showpacs,amssymb,epsf]{revtex4}
\usepackage{graphicx}
\usepackage{bm}

\usepackage{amsmath}

\begin{document}

\title{Effective contact model for transport through
weakly-doped graphene}

\author{Henning Schomerus}
\affiliation{Department of Physics, Lancaster University,
Lancaster, LA1 4YB, UK}

\date{\today}

\begin{abstract}

Recent investigations address  transport through ballistic
charge-neutral graphene strips coupled to doped graphitic leads.
This paper shows that identical transport properties arise when
the leads are replaced by quantum wires. This duality between
graphitic and metallic leads originates in the selection of modes
with transverse momentum close to the K points, and can be
extended to a wide class of contact models. Among this class we
identify a simple, effective contact model, which provides an
efficient tool to study the transport through extended
weakly-doped graphitic systems.

\end{abstract}
\maketitle

\section{Introduction}

Since the recent break-through in its fabrication,
\cite{Novoselov2004,Novoselov2005,Zhang2005,Zhang2006} graphene,
the atomically thin two-dimensional hexagonal arrangement of
carbon atoms, has caught the excitement of experimentalists and
theoreticians alike because it possesses unique electronic
properties which originate from the two conical points of its
Fermi surface, located at the K points of the hexagonal Brillouin
zone. In a simple model, a clean graphene sheet is described by a
tight-binding
%HS1: hamiltonian
Hamiltonian $H=-\gamma\sum_{\langle
ij\rangle}c_i^\dagger c_j$, in which next neighbors $\langle
ij\rangle$ on the hexagonal lattice (with lattice constant $a$)
are connected by a hopping matrix element $\gamma$. The lattice
supports Bloch waves with wave numbers $k_x$, $k_y$ and dispersion
relation $E=\pm \gamma|1+2e^{i3k_xa/2}\cos({\sqrt{3}k_ya /2})|$.
At the Fermi energy $E=0$ of undoped, charge neutral graphene, the
Fermi surface shrinks to two inequivalent points
$(k_x,k_y)=(\frac{2\pi}{3a},\pm\frac{2}{3}\frac{\pi}{\sqrt{3}a})$.
 In the vicinity of these points the dispersion relation can be linearized, and then becomes
conical with slope $|dE/dk|=3\gamma a/2$. The ensuing low-energy
theory is described by a Dirac Hamiltonian. \cite{dirac}

By varying the Fermi energy via a gate voltage across the
charge-neutrality point, graphene offers the unique possibility to
switch the doping of the system from n to p, where the doped
charge density can be changed continuously. The first
graphene-based field-effect transistor was demonstrated in Ref.\
\onlinecite{Novoselov2004}. The mobility of carefully fabricated
graphene flakes already exceeds 50,000 cm$^2$/Vs. \cite{Zhang2006}
 Motivated by the ensuing prospect of graphitic
electronic devices with properties very much different from
semiconductor-based technology, a number of recent works have
explored the phase-coherent transport properties of finite
segments of graphene, connected via leads to electronic
reservoirs. This resulted in the discovery of a modified quantum
Hall effect \cite{Novoselov2005,Zhang2005} with a half-integer
sequence of Hall conductance plateaus. It was furthermore
predicted that weak-localization corrections to the conductance
depend sensitively on symmetries preserved or broken by the
disorder. \cite{mccann,Morozov} Clean graphitic samples exhibit a
finite conductivity  of the order of the conductance quantum
\cite{Novoselov2004,Novoselov2005,Zhang2005,tworzydlo,guinea,theories}
and a shot noise identical to a disordered quantum wire.
\cite{tworzydlo} Transport across p-n junctions was theoretically
studied in Ref.\ \onlinecite{cheianov}. A similar but modified set
of unique transport phenomena is found for graphene bilayers,
which furthermore offer the possibility to open up a controllable
gap by an electric field effect (see Ref.\ \onlinecite{mccanngap}
and references therein).

A problem untouched by the recent transport investigations is the
role of the leads connecting the graphitic sample to the
electronic reservoirs. Some of the recent theoretical transport
studies model the leads as strips of doped graphene. This type of
lead supports distinctively different sets of modes than a
conventional quantum wire
--- for the same transverse mode profile, propagating modes in a
conventional quantum wire often correspond to evanescent modes in
a graphene strip, and vice versa. Therefore, it could be argued
that a detailed
%HS modeling
modelling of the leads is critical for the
understanding of transport in graphitic systems.

The objective of this paper is to show that to the contrary,
transport through a sufficiently large and only weakly doped
graphene sample often does {\rm not} critically depend on most
details of the leads. For the particular case of quantum-wire and
graphitic leads, this insensitivity manifests itself in an
explicit duality, which can be formulated in terms of a shift of
the gate potentials that control the charge density in the leads.
This duality originates in a mode selection mechanism, which
dictates that the transport in extended, only weakly doped samples
of graphene is dominated by the small part of the mode space in
the vicinity of the conical points. Using this universal
mechanism, one is naturally led to formulate an effective contact
model, which is parameterized by a single complex number $\mu$.
This effective model can be applied to a much larger class of
leads which are amenable to the mode-selection mechanism. The main
requirement is that the leads are sufficiently wide, and provide a
dense set of propagating modes with transverse wave number close
to the conical points (leads which do not possess any propagating
modes at these transverse wave numbers exhibit a very large
contact resistance, and hence do not provide a good electronic
coupling).

In striking contrast to universal transport through semiconductor
quantum dots \cite{beenakkerreview}, the effective contact model
provides a unified description of leads which support a different
{\em number} of propagating modes (but share the same width).
Numerical computations confirm that the effective contact model
provides an efficient and reliable tool to study the transport
through extended weakly-doped graphitic systems.

This paper is organized as follows. Section \ref{sec2a} formally
poses the problem of the lead-sensitivity of transport in
mesoscopic systems. In Section \ref{sec2}, I calculate the
conductance and Fano factor of a rectangular graphene sample which
is connected to quantum wires or graphitic leads. The rectangular
graphene sample is held at the charge neutrality point, while the
charge density in the leads is controlled by gate potentials. At a
finite charge density in the leads, this deliver the duality of
both contact models. In Section \ref{sec3}, I describe the
effective contact model which represents a large class of leads,
encompassing  quantum wires and  doped graphitic strips. Section
\ref{sec4} describes how the complex number $\mu$ could be
extracted from the conductance and shot noise, and tests the
applicability of the effective contact model via numerical
computations for rectangular and circular graphene samples.
Conclusions are presented in \ref{sec5}. The appendices contain
technical details of the calculations.

\begin{figure}[t]
\begin{center}
\includegraphics[width=\columnwidth]{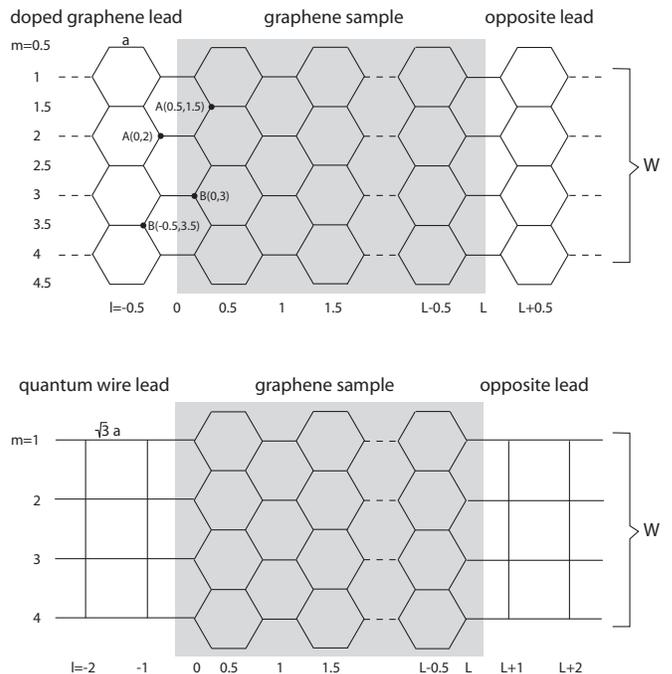}
\end{center} \caption{A graphene sample, modelled
as a hexagonal lattice (lattice constant $a$), is attached to
leads  formed either by doped graphene (top)  or by a quantum
wire, modelled as a commensurably matched square lattice with
lattice constant $\sqrt{3} a$ (bottom). The charge carrier density
in the leads is controlled by a gate potential $V_g$, while the
central graphitic region is held close to the charge-neutrality
point. The paper demonstrates that both leads are equivalent when
the gate potential is suitably adjusted, and further extends this
duality to a broad class of leads and contacts, which all share
the properties of a simple effective contact model.
 \label{fig1}}
\end{figure}

\section{\label{sec2a}Formulation of the problem}

At low bias, the %HS2 hyphenated
phase-coherent transport properties of a
mesoscopic system probed in  a two-terminal geometry are encoded
in the scattering matrix \cite{beenakkerreview,blanterreview}
\begin{equation}
S=\left(%
\begin{array}{cc}
  r & t' \\
  t & r' \\
\end{array}%
\right),
\end{equation}
which contains the transmission  (reflection) amplitudes $t$, $t'$
($r$, $r'$) of charge carriers incident from the source or drain
contact, respectively. Two characteristic transport properties
 are the Landauer conductance
\begin{equation}\label{eq:gdef}
G=(2e^2/h)\,{\rm tr}\, t^\dagger t
\end{equation}
and the shot-noise Fano factor
\begin{equation}\label{eq:fdef}
F=1-\frac{{\rm tr}\, t^\dagger t t^\dagger t}{\,{\rm tr}\,
t^\dagger t}.
\end{equation}
In general, the transmission matrix $t$ is determined by matching
the propagating modes in the leads to the modes in the mesoscopic
system, and hence depends on the properties of both of these
constituents. This paper considers the case that the mesoscopic
system is an extended, only weakly doped sample of graphene with a
specified geometry, and asks the questions how the transport
properties change when the sample is contacted by different types
of leads (keeping the width and position of the leads fixed).

\section{\label{sec2}Duality of metallic and graphitic contacts}

It is instructive to first explore the potential relevance and
eventual insensitivity to most details of the leads by calculating
the conductance for a specific example, a rectangular undoped
graphene sample of width ${\cal W}=\sqrt{3} a W $ and length
${\cal L}=3 a L$, where $W, L \gg 1$ are integers (see  Fig.\
\ref{fig1}). The strip is connected to leads of the same width,
either formed by doped graphene or by a quantum wire, which are
modelled in a tight-binding approach on a hexagonal or square
lattice, respectively. The electronic density in the leads is
controlled by a gate potential, inducing an on-site potential
energy denoted by $V_g$. For $\gamma/W\ll |V_g|\ll \gamma$ the
transport of this system connected to graphitic leads has been
investigated earlier in the framework of the Dirac equation.
\cite{tworzydlo} It was found that the conductivity
$\sigma=\frac{\cal L}{\cal W}G=\frac{4e^2}{\pi h}$ is of order of
the experimentally observed value, while the Fano factor $F=1/3$
coincides with the universal value of a disordered quantum wire.

A unified description of square- and hexagonal-lattice leads can
be achieved when the hexagonal sublattices $A$, $B$, which differ
in the orientation of the bonds, are indexed by two numbers $l$,
$m$ which are either both integer or both half-integer (see Fig.\
\ref{fig1}). On the square lattice the indices $l$ and $m$ are
both integer. For both types of lattice, the transverse wave
function of the modes in the lead are then given by
\begin{equation}\label{eq:psi}
\Psi_{nm}= \sqrt{\frac{2}{W+1}}\sin\frac{nm\pi}{W+1},
\end{equation}
 where
$n=1,2,3,\ldots,W$ is the mode index. Associated with each
transverse mode are two extended Bloch waves with longitudinal
wave numbers $\pm k_n$, which are real for propagating modes and
complex for evanescent modes. The wave numbers are fixed by the
dispersion relation, which for the hexagonal lattice is given by
\begin{subequations}\label{hdisp}
\begin{eqnarray}
&&V_g=\eta\sqrt{f_{n,k_n}f_{n,-k_n}},\quad \eta={\rm sgn}\,V_g=\pm 1,\\
&&f_{n,k_n}=\gamma+2 \gamma e^{i3k_na/2}\cos\frac{n\pi}{2(W+1)},
\end{eqnarray}
\end{subequations}
while for the square lattice
\begin{eqnarray}\label{sdisp}
&& V_g= 2\gamma \cos(\sqrt{3}ak_n)+2\gamma \cos\frac{n\pi}{W+1} .
\end{eqnarray}

Because of the different dispersion relations, the sets of
transverse-mode indices $n$ supporting propagating modes on the
two types of lattice are not the same. Hence, for ordinary
mesoscopic systems (such as a semiconducting quantum dot, a
quantum wire, a metallic nanoparticle or a hybrid structure with
superconducting or ferromagnetic properties) one would obtain
distinctively different transport properties when quantum wire
leads would be replaced by graphitic leads.

\begin{figure}[t]
\begin{center}
\includegraphics[width=\columnwidth]{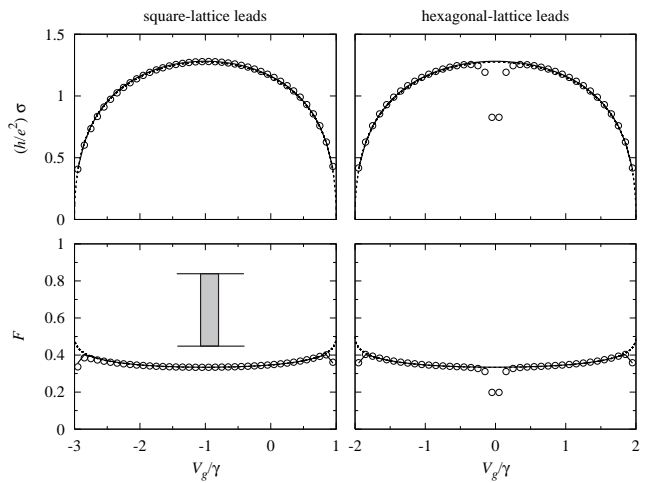}
\end{center} \caption{Gate-voltage dependence of the conductivity $\sigma=({\cal L}/{\cal W}) G$ (in units of
$e^2/h$) and the shot-noise Fano factor $F$ of an undoped graphene
strip of width ${\cal W}=152\,\sqrt{3}\,a$ and length ${\cal
L}=30\,a$. The open data points are obtained by numerical
computations for square-lattice  leads (left) and
hexagonal-lattice leads (right). The solid curves are the
prediction of the effective contact model (\ref{result}) with
$\mu$ given by Eq.\ (\ref{muh}) or Eq.\ (\ref{mus}), respectively.
The dashed curves are the analytical expressions
(\ref{eq:gg},\ref{eq:gs},\ref{eq:fg},\ref{eq:fs}) obtained within
the saddle-point approximation.
 \label{fig2}}
\end{figure}

In order to obtain the transmission amplitudes $t_{n'n}$ when
these leads are coupled to the undoped rectangular graphene
sample, the modes in the leads have to be matched to modes in the
graphene strip, which follow from Eq.\ (\ref{hdisp}) by setting
$V_g=0$. For the present quasi-one dimensional geometry, the mode
index $n$ is conserved, \cite{tworzydlo} and a straightforward
calculation (see Appendix \ref{app:a}) delivers
\begin{equation}\label{eq:tnn}
t_{n'n}=\delta_{n'n}
\frac{\mu_{n,+}-\mu_{n,-}}{\mu_{n,+}e^{-\kappa_n{\cal
L}}-\mu_{n,-}e^{\kappa_n{\cal L}}},
\end{equation}
where the moduli of
\begin{equation}\label{eq:kappa}
\kappa_n =\frac{2}{3a} \ln \left(2\cos\frac{n\pi}{2(W+1)}\right)
\end{equation}
 denote the
decay constants of the modes in undoped graphene. The lead
dependence of the transmission coefficients is encoded in the self
energies $\mu_{n,\pm}$ of the in- and outgoing modes, given by
\begin{subequations}\label{mu}
\begin{eqnarray}  \mu^{(h)}_{n,\pm}= &-\eta\gamma
\frac{f_{n,\pm \eta k_n}}{\sqrt{f_{n,k_n}f_{n,-k_n}}} &
\mbox{(hexagonal lattice)},
\\
 \mu^{(s)}_{n,\pm}=& -\gamma e^{\pm i\sqrt{3}k_n a}\qquad  &
\mbox{(square lattice)}.
\end{eqnarray}
\end{subequations}
The effective contact model developed in Section \ref{sec3} arises
from the observation that the detailed lead dependence embodied in
these numbers  becomes irrelevant when the leads are coupled to a
sufficiently large graphitic region. According to Eq.\
(\ref{eq:kappa}),
  all modes  in this region decay rapidly with the exception of modes with index $n \approx
2(W+1)/3\equiv \tilde n$, which have transverse wave numbers in
the vicinity of the conical points. For a sufficiently large
sample we hence only require to know the complex number
$\mu\equiv\mu_{\tilde n,+}=\mu_{\tilde n,-}^*$, which for a
hexagonal-lattice lead is given by
\begin{eqnarray}
\label{muh}
\mu^{(h)}(V_g)=-\frac{1}{2}\left(V_g+i\sqrt{4\gamma^2-V_g^2}\right),
\end{eqnarray}
while for a square-lattice lead
\begin{eqnarray}\label{mus}
\mu^{(s)}(V_g)=
-\frac{1}{2}\left(V_g+\gamma+i\sqrt{4\gamma^2-(V_g+\gamma)^2}\right).
\end{eqnarray}

Before we describe the general consequences of this observation we
first proceed to explore the consequences for the rectangular
graphene sample. To make contact to  Ref.\ \onlinecite{tworzydlo},
let us further assume that the strip is very wide, ${\cal W}\gg
{\cal L}\gg a$. The conductance
 can then be calculated in a saddle-point approximation, which is described  in Appendix \ref{app:b}. For
 hexagonal-lattice
leads this gives
\begin{eqnarray}\label{eq:gg}
G&=&\frac{4 e^2}{\pi h}\frac{{\cal W}}{{\cal L}
}\frac{\sqrt{4\gamma^2-V_g^2}}{V_g}\arcsin\frac{V_g}{2\gamma},
\end{eqnarray}
which for $V_g\to 0$ recovers the result $G= \frac{4 e^2}{\pi
h}\frac{{\cal W}}{{\cal L} }$  derived from the Dirac equation.
\cite{tworzydlo} For square-lattice leads one finds
\begin{eqnarray}\label{eq:gs}
G&=&\frac{4 e^2}{\pi h}\frac{{\cal W}}{{\cal L}
}\frac{\sqrt{4\gamma^2-(V_g+\gamma)^2}}{V_g+\gamma}\arcsin\frac{V_g+\gamma}{2\gamma}.
\end{eqnarray}
A similar calculation yields the Fano factor
\begin{eqnarray}\label{eq:fg}
F&=&
\frac{2\gamma^2}{V_g^2}-\frac{\sqrt{4\gamma^2-V_g^2}}{2V_g\arcsin(V_g/2\gamma)}
\end{eqnarray}
for the hexagonal-lattice leads, and
\begin{eqnarray}
 \label{eq:fs} F&=&
\frac{2\gamma^2}{(V_g+\gamma)^2}-\frac{\sqrt{4\gamma^2-(V_g+\gamma)^2}}{2(V_g+\gamma)
\arcsin[(V_g+\gamma)/2\gamma]}
\end{eqnarray}
for the square-lattice leads. For a numerical validation of
these results see Section \ref{sec4}.

Equations (\ref{eq:gg},\ref{eq:fg}) for the hexagonal-lattice
leads and (\ref{eq:gs},\ref{eq:fs}) for the square-lattice leads
coincide when the gate potential is shifted by $\gamma$. In
particular, for the square-lattice leads the values  $G= \frac{4
e^2}{\pi h}\frac{{\cal W}}{{\cal L} }$ of the conductance  and
$F=1/3$ for the Fano factor are now recovered for  $V_g\to
-\gamma$. This is a direct consequence of the relation
\begin{eqnarray} \label{eq:relation}
\mu^{(s)}(V_g)=\mu^{(h)}(V_g+\gamma)
\end{eqnarray}
between the characteristic self energies $\mu$, Eqs.\ (\ref{muh})
and (\ref{mus}), which describe the coupling
 of the leads  to the conical points.
This relation does {\em not} hold for the self energies (\ref{mu})
away from the conical point, but this is not detected in  the
transport as long as the leads support sufficiently many
propagating modes around $n=\tilde n$.
 Surprisingly, Eq.\
(\ref{eq:relation}) entails that the point of half-filling of the
hexagonal-lattice leads (the charge-neutrality point $V_g=0$)
corresponds to the point of three-quarter filling in the
square-lattice leads.

\section{\label{sec3}Effective contact model}

In order to elucidate the generality of this duality between
hexagonal- and square-lattice leads, we now liberate ourselves
from the strip geometry of the graphene sample, hence, consider
samples of more arbitrary geometry in which different transverse
modes can be mixed by the transport. In this case the matrix of
transmission amplitudes $t_{n'n}$ is no longer diagonal, and in
general can be obtained from the Fisher-Lee relation
\cite{fisherlee}
\begin{equation}
S=-1+ i\sqrt{v} \Psi^\dagger {\cal P}^\dagger (E-H-
\Sigma)^{-1}{\cal P} \Psi \sqrt{v}.
\end{equation}
The leads are represented by their self energy $\Sigma= {\cal P}
\Psi\,{\rm diag}\, (\mu_{n,+})\Psi^\dagger {\cal P}^\dagger$,
where ${\cal P}$ is a  coupling matrix of the leads to the contact
region. The diagonal matrix $v={\rm diag}\,(-2\,{\rm Im}\,
\mu_{n,+})$ contains factors proportional to the propagation
velocity of the modes in the leads. The matrix $\Psi$ now accounts
for the transverse modes in all leads.

According to Eq.\ (\ref{mu}), the modes in the hexagonal- and
square-lattice leads in general have different propagation
velocities and self-energies. As a consequence, when these leads
are connected to an arbitrary system, described by the internal
Hamiltonian $H$, the resulting transport properties in general
will differ. When $H$ represents a sufficiently large, weakly
doped graphitic system, however, the localization of the Fermi
surface near the conical points guarantees that only the lead
modes coupling to this part of the Brillouin zone will contribute
significantly to the transport. This also applies to other types
of leads, different from the two specific cases considered so far,
and then yields the universality among contact models advertised
in the introduction.

Assuming that the leads support a large number of propagating
modes whose properties depend smoothly on the quasi-continuous
mode index $n$, we hence can equip all modes in the lead with the
same constant $\mu_{n,+}\to \mu\equiv\mu_{\tilde n,+}$
characteristic of the modes coupling to the conical points. Using
that the transverse mode profiles form an orthogonal set, $\Psi
\Psi^\dagger=1$, the self energy  then simplifies to $\Sigma= \mu
{\cal P} {\cal P}^\dagger$. For optimally matched leads, ${\cal
P}=P$ is just a projector of the internal system space onto the
contact region. A unitary transformation $S\to \Psi S
\Psi^\dagger$ (which does not affect the conductance and the Fano
factor) then results in a simplified Fisher-Lee relation for the
scattering matrix,
\begin{equation} \label{result}
S=-1- (2 i \,{\rm Im}\, \mu) P^T (E-H-\mu PP^T)^{-1}P.
\end{equation}
This form of the scattering matrix defines an effective contact
model, which is parameterized by a single complex number $\mu$.

For hexagonal-lattice or square-lattice leads, $\mu$ is given by
Eq.\ (\ref{muh}) or Eq.\ (\ref{mus}), respectively. However, our
argumentation entails that the simple effective model Eq.
(\ref{result}) can be applied to a much larger class of leads,
which share the same mode-selection principle when coupled to
graphene. (An example would be a square lattice with a reduced
lattice constant $\sqrt{3}a/q$, where $q$ is an integer.) Within
this class of leads, $\mu$ is still given by the self energy
$\mu_{\tilde n,+}$. Depending on the contacts, the matrix ${\cal
P} {\cal P}^\dagger$ may not be uniform over the contact region,
and also may account for a contact tunnel-barrier resistance.
Assuming that the coupling strengths $p_n= \langle \Psi_n^\dagger
{\cal P} {\cal P}^\dagger \Psi_n\rangle/\langle \Psi_n^\dagger P
P^T \Psi_n\rangle$ (where $\Psi_n$ is the column vector associated
to the $n$th transverse mode) do not depend strongly on the mode
index, the effective contact model (\ref{result}) still holds with
$\mu=p_{\tilde n}\mu_{\tilde n,+}$. \cite{remark2} Ideal coupling
into the graphitic modes near the conical points is realized for a
self energy $\mu=-i\gamma$.

\section{\label{sec4}Applications}

This section describes some consequences of the effective contact
model (\ref{result}), and compares its predictions to the results
of numerical computations.

\subsection{Transport in a wide, long graphitic
strip and the reconstruction of $\mu$}

Within the effective contact model (\ref{result}), the
conductance of a wide and long graphene strip (as considered in
Section \ref{sec2}) is given by
\begin{eqnarray}\label{eq:ggeneral}
G&=&\frac{4e^2}{\pi h}\frac{{\cal W}}{{\cal L}}\frac{{\rm
Im}\,\mu}{{\rm Re}\,\mu}\arcsin\frac{{\rm Re}\,\mu}{|\mu|},
\end{eqnarray}
while the Fano factor is given by
\begin{eqnarray}\label{eq:fgeneral}
F=\frac{|\mu|^2}{2({\rm Re}\,\mu)^2}-\frac{{\rm Im}\,\mu}{2{\rm
Re}\,\mu \arcsin({\rm Re}\,\mu/|\mu|)}
\end{eqnarray}
(for the derivation see Appendix \ref{app:b}). These formulae
generalize Eqs. (\ref{eq:gg}) -- (\ref{eq:fs}),
%HS3 (\ref{eq:gg}), (\ref{eq:gs}), (\ref{eq:fg}), (\ref{eq:fs}),
which are recovered when $\mu$ is taken from Eq. (\ref{muh}) (for
hexagonal-lattice leads)
%HS4 or Eq. (\ref{muh}) (for hexagonal-lattice leads).
or Eq. (\ref{mus}) (for square-lattice leads). For other types of
leads, the two transport characteristics deliver two independent
numbers which can be used to infer the complex number $\mu$.

\subsection{Numerical results}

\begin{figure}[t]
\begin{center}
\includegraphics[width=\columnwidth]{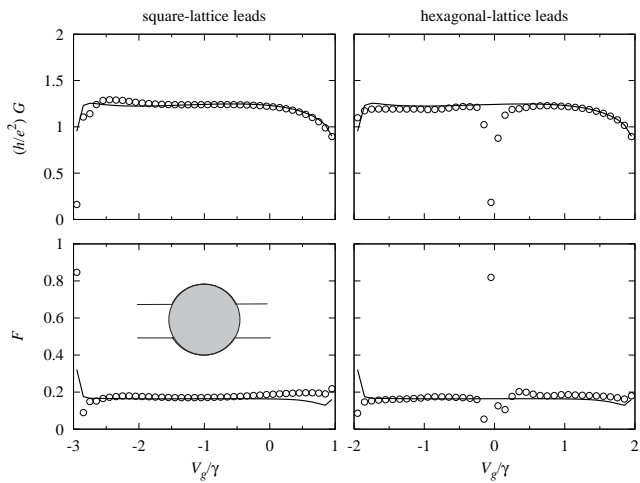}
\end{center} \caption{Gate-voltage dependence of the conductance $G$ (in units of $e^2/h$) and the shot-noise Fano
factor $F$  for a weakly doped circular graphene sample
($E_F=\gamma/10$) of radius ${\cal R}= 100\,\sqrt{3}\,a$,
connected to leads of width ${\cal W}= 60\,\sqrt{3}\,a$. The data
points are obtained by numerical computations for square-lattice
leads (left) and hexagonal-lattice leads (right). The curves are
the predictions of the effective contact model (\ref{result}) with
$\mu$ given by Eq.\ (\ref{muh}) or Eq.\ (\ref{mus}), respectively.
 \label{fig3}}
\end{figure}

The validity of the effective contact model can be asserted by
numerical computations. The gate-voltage dependence of the
conductivity $\sigma$ and the Fano factor $F$ for a sample of
width ${\cal W}=152\,\sqrt{3}\,a$ and length ${\cal L}=30 \,a$
is shown in Fig.\ \ref{fig2}. Results of numerical computations
obtained by the method of recursive Green's functions
\cite{decimation} are compared to the theoretical prediction of
the effective contact model resulting from Eq.\ (\ref{muh}) or
Eq.\ (\ref{mus}), as well as to the analytical expressions
%HS5 (\ref{eq:gg},\ref{eq:gs},\ref{eq:fg},\ref{eq:fs}).
(\ref{eq:gg}) -- (\ref{eq:fs}). Good agreement is found in the
range $|V_g+\gamma|< 2\gamma$ for the square-lattice leads, and
$\gamma/W\lesssim |V_g|< 2\gamma$ for the hexagonal-lattice leads,
corresponding to the condition that the modes in the vicinity of
$\tilde n$ are propagating.

As an additional example, Fig.\ \ref{fig3} shows the conductance
and the Fano factor of a weakly doped circular graphitic region
($E_F=0.1\,\gamma$), calculated for hexagonal- and square-lattice
leads and within the effective contact model (\ref{result}), where
$\mu$ is again given by Eqs. (\ref{muh}), (\ref{mus}).
%HS6 $\mu$ is  given by Eqs. (\ref{muh},\ref{mus}).
With the exception of the region $|V_g|\approx \gamma/W$, where
the hexagonal-lattice lead does not support many propagating
modes, the predictions of the effective contact model again agree
nicely with the results for the two types of leads. This shows
that the effective contact model remains applicable for graphitic
samples where modes are mixed in the transport (so that the
transmission matrix is no longer diagonal).

\section{\label{sec5}Summary}

This paper critically assesses how sensitively the transport
through extended samples of weakly doped graphene depends on the
details of the leads and contacts connecting the sample to the
electronic reservoirs. The starting point is the observation that
there exists a duality between doped graphitic leads and quantum
wires, which result in the same transport properties if  a gate
voltage is suitably adjusted. This duality is {\em not} based on
similarities of the two type of wires (which would result in very
distinct transport properties if attached to a conventional
material), but is rooted in a unique mode selection mechanism
which originates from the conical points of undoped graphene.
Maybe most strikingly, the duality holds even though both types of
leads support a different {\em number} of propagating modes.

Since the mode selection mechanism is a universal property of
weakly doped graphene, the duality described above can be
generalized into an effective contact model, which
%%HS7 describes
applies to a broad class of leads. The effective model is
parameterized by a single complex number $\mu$, which can be
%HS8 determeind
determined by a measurement of the conductance and the shot noise
of a rectangular undoped graphene strip.

In the present paper, this effective model has been derived using
the technical requirement that the leads provide a densely spaced
set of propagating modes which couple smoothly to the conical
points. This requirement is fulfilled for the typical leads
considered in past and present mesoscopic-transport studies, and
especially, for ballistic, wide contacts which provide a good
electronic coupling to the graphitic sample (so that a transport
measurement probes the sample, and not merely the contact
resistance).

One potential mechanism to violate the specific technical
assumptions used to derive the effective contact model is
strong interface disorder, so that the propagating modes in the
leads are mixed by the interface. For an ordinary mesoscopic
system, such a diffusive interface changes the specific
transport properties decisively (these changes cannot be
described by a simple tunnel barrier). For weakly doped
graphene, it is conceivable that even such a drastic
modification of the contacts does not completely violate the
effective contact model. In a transfer description of the
contact, the crossing of the diffusive interface from the
graphitic sample to the lead results in a translation of the
transverse mode profile of the modes close to the conical
points into a random superposition of the modes in the
% HS9 leads.
lead. It is conceivable that self-averaging in these random
superpositions merely renormalizes the effective self-energy $\mu$
into the average self-energy of the propagating modes in the lead.
This question goes beyond the scope of the techniques used in the
present paper, and is left for further consideration.

%HS10 added:
\acknowledgements

I gratefully acknowledge helpful discussions with Carlo Beenakker,
Edward McCann, and
 John P. Robinson.
 This work was supported
by the European Commission, Marie Curie Excellence Grant
MEXT-CT-2005-023778.

\appendix
\section{\label{app:a}Mode matching for an undoped rectangular graphene strip}

\begin{figure}[t]
\begin{center}
\includegraphics[width=\columnwidth]{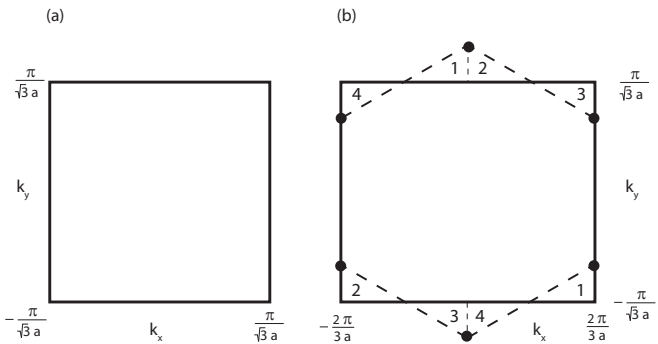}
\end{center} \caption{(a) Brillouin
%HS11 zones
zone for a square lattice with lattice constant $\sqrt{3}\,a$. (b)
Hexagonal Brillouin zone for a honeycomb lattice with lattice
constant $a$ (dashed), and its %HS rearrangment
rearrangement into a rectangular
Brillouin zones (solid). The numbered triangles denote regions
which are transported with reciprocal lattice vectors. The solid
dots denote the K points.
 \label{fig4}}
\end{figure}

This appendix provides details of the derivation of the
transmission eigenvalues (\ref{eq:tnn}) for an undoped
rectangular graphitic strip coupled to square-lattice or
hexagonal-lattice leads.

A unified description of both cases in real space is
facilitated by the discrete coordinate system $(l,m)$, shown in
Fig. \ref{fig1}. Analogously, Fig. \ref{fig4} shows a
convenient choice of the Brillouin zones which facilitates a
unified description in $k$ space. For the square lattice with
lattice constant $\sqrt{3}a$, the Brillouin zone is given by
the standard square $|k_x|,|k_y|\leq \frac{\pi}{\sqrt{3}a}$.
Using the periodicity of the band structure, the conventional
hexagonal Brillouin zone for the honeycomb lattice is
rearranged into the rectangle $|k_x|\leq \frac{2\pi}{3a}\equiv
q_x$, $|k_y|\leq \frac{\pi}{\sqrt{3}a}\equiv \frac{3}{2}q_y$,
where $|k_x|=q_x$, $|k_y|=q_y$ denotes the position of the
conical points. With this choice, both Brillouin zones have the
same extent in $k_y$ direction. Moreover, the boundary
conditions of the tight-binding model select the same
transverse wavenumbers $k_{y,n}=\frac{n\pi}{{\cal W}+\sqrt{3}
a}$. The transverse mode profile $\Psi_{nm}$ is then uniform
throughout the system, and given by Eq.\ (\ref{eq:psi}).

For fixed transverse wavenumber $k_{y,n}$, the longitudinal
wavenumbers $k_n$ follow from the dispersion relations
(\ref{hdisp}) and (\ref{sdisp}), respectively. For each
transverse wavenumber, there are two solutions of opposite
sign.

In the undoped central graphene regions, the longitudinal
wavenumbers are complex, and the modes are evanescent with decay
constant ${\rm Im}\,k_n=\kappa_n$ given in Eq.\ (\ref{eq:kappa}).
The wave functions $\phi_A$, $\phi_B$ on the $A$ and $B$ sites are
determined by the Schr{\"o}dinger equation, which delivers
\begin{subequations}\label{eq:psigraph0}
\begin{eqnarray}
&&\phi_A^{(n)}(l,m)=\alpha e^{-3 a l\kappa_n+2\pi i l}\Psi_{nm},
\\
&&\phi_B^{(n)}(l,m)=\beta e^{3 a l\kappa_n+2\pi i l}\Psi_{nm},
\end{eqnarray}
\end{subequations}
respectively, where $\alpha$ and $\beta$ are
constants.

The Landauer approach requires to match these wavefunctions to the
propagating modes in the lead. These modes are characterized by a
real longitudinal wavenumbers $\pm k_n$, where for definiteness
$k_n>0$ is taken to be the solution in the right half of the
Brillouin zone.

On the hexagonal lattice, modes with a positive longitudinal
wavenumber $k_n>0$ are right-propagating when ${\rm sgn}
V_g=\eta>0$, (so that the Fermi energy lies below the conical
point), while these modes are left-propagating for $\eta <0$
(since the band structure above the conical point has the
opposite curvature). For a particle incident from the left, the
solution of the Schr{\"o}dinger equation in the left lead takes
the form
\begin{subequations}\label{waves}
\begin{eqnarray}
&&\phi_{A,L}^{(n)}(l,m)=C \eta (f_{n,-\eta k_n}/f_{n,\eta
k_n})^{1/4} e^{3i\eta k_n a l}\Psi_{nm}\nonumber \\ &&{}\quad +C
r_{nn} \eta (f_{n,-\eta k_n}/f_{n,-\eta k_n})^{1/4} e^{-3i\eta k_n
a l}\Psi_{nm},
\\
&&\phi_{B,L}^{(n)}(l,m)=C  (f_{n,\eta k_n}/f_{n,-\eta k_n})^{1/4}
e^{3i\eta k_n a l}\Psi_{nm}\nonumber \\ &&{}\quad +C r_{nn}
(f_{n,-\eta k_n}/f_{n,\eta k_n})^{1/4} e^{-3i\eta k_n a
l}\Psi_{nm},
\end{eqnarray}
while in the right lead it is given by
\begin{eqnarray}
&&\phi_{A,L}^{(n)}(l,m)=C t_{nn} \eta (f_{n,-\eta k_n}/f_{n,\eta
k_n})^{1/4} e^{3i\eta k_n a (l-L)}\Psi_{nm},\nonumber \\
\\
&&\phi_{B,L}^{(n)}(l,m)=C  t_{nn} (f_{n,\eta k_n}/f_{n,-\eta
k_n})^{1/4} e^{3i\eta k_n a (l-L)}\Psi_{nm}.\nonumber \\
\end{eqnarray}
\end{subequations}
Here $C$ is an arbitrary coefficient, while $r_{nn}$ and $t_{nn}$
are the reflection and transmission coefficient, respectively.

On the square lattice,  positive longitudinal wavenumbers
$k_n>0$ are always associated with  right-propagating modes.
The wave function in the left lead can hence be written as
\begin{subequations}\label{eq:psis}
\begin{eqnarray}
\phi_{L}^{(n)}(l,m)&=& C e^{\sqrt{3}ik_n a(l+1/2)}\Psi_{nm}
\nonumber \\ &&{}+ C r_{nn} e^{- \sqrt{3}ik_n a(l+1/2)}\Psi_{nm},
\end{eqnarray}
while in the right lead
\begin{eqnarray}
&&\phi_{R}^{(n)}(l,m)= C t_{nn} e^{\sqrt{3}ik_n
a(l-L-1/2)}\Psi_{nm}.
\end{eqnarray}
\end{subequations}

Using the fact that the wavefunctions (\ref{eq:psigraph0}),
(\ref{waves}), (\ref{eq:psis}), all satisfy the Schr{\"o}dinger
equation when the graphitic strip or the leads would be formally
extended beyond the interface, the matching conditions take the
simple form of continuity requirements
\begin{subequations}\label{eq:cont1}
\begin{eqnarray}
&& \phi_A^{(n)}(0,m)=\phi_{A,L}^{(n)}(0,m),
\\
&&\phi_B^{(n)}(0,m)=\phi_{B,L}^{(n)}(0,m),
\\
&& \phi_A^{(n)}(L,m)=\phi_{A,R}^{(n)}(L,m),
\\
&&\phi_B^{(n)}(L,m)=\phi_{B,R}^{(n)}(L,m)
\end{eqnarray}
\end{subequations}
for the hexagonal-lattice leads, and
\begin{subequations}\label{eq:cont2}
\begin{eqnarray}
&& \phi_A^{(n)}(0,m)=\phi_{L}^{(n)}(-1,m),
\\
&&\phi_B^{(n)}(0,m)=\phi_{L}^{(n)}(0,m),
\\
&& \phi_A^{(n)}(L,m)=\phi_{R}^{(n)}(L,m),
\\
&&\phi_B^{(n)}(L,m)=\phi_{R}^{(n)}(L+1,m)
\end{eqnarray}
\end{subequations}
for the square-lattice leads.

The resulting linear systems of equations deliver the transmission
coefficient $t_{nn}$ in the form (\ref{eq:tnn}), where for each
lattice the self energy is defined in Eq. (\ref{mu}).

\section{\label{app:b}Conductance and Fano factor for wide and long graphene strips}

This appendix describes the saddle-point approximation which is
used to obtain the conductance (\ref{eq:ggeneral}) and the Fano
factor (\ref{eq:fgeneral}) of an undoped graphene strip of
width ${\cal W}\gg {\cal L}\gg a$. The conductance
(\ref{eq:gg}), (\ref{eq:gs}) and the Fano factor (\ref{eq:fg}),
(\ref{eq:fs}) for hexagonal-lattice and square-lattice leads
follow when these expressions are evaluated with (\ref{muh})
and (\ref{mus}), respectively.

Let us start with a detailed account for the conductance, which
in general is related to the transmission coefficients $t_{nm}$
via the Landauer formula (\ref{eq:gdef}). The transmission
matrix (\ref{eq:tnn}) of the graphene strip is diagonal, hence
$G=(2e^2/h)\sum_n|t_{nn}|^2$. Since the strip is wide, ${\cal
W}\gg  {\cal L}\gg a$, the transverse wavenumbers are closely
spaced and the transmission coefficients (\ref{eq:tnn}) depend
quasi-continuously on the mode index $n$, so that the sum can
be replaced by an integral,
\begin{equation}\label{eq:gint}
G=\frac{2e^2}{h}\int\left|\frac{\mu_{n,+}-\mu_{n,-}}{\mu_{n,+}e^{-\kappa_n{\cal
L}}-\mu_{n,-}e^{\kappa_n{\cal L}}}\right|^2\,dn.
\end{equation}
For ${\cal L}\gg a$, all transmission coefficients decay rapidly
with the exception of modes with index $n \approx 2(W+1)/3\equiv
\tilde n$, which have transverse wave numbers in the vicinity of
the conical points and hence a small decay coefficient $\kappa_n$,
given in  Eq.\ (\ref{eq:kappa}). Hence the integrand in Eq.
(\ref{eq:gint}) has a pronounced maximum at $n \approx \tilde n$.
Away from this maximum, for $n = \tilde n+\delta$, the integrand
decays rapidly because of the finite decay constant $\kappa_n
\approx \delta \pi/{\cal W}$, and attains very small values on a
scale on which the self-energies $\mu_{n,+}=\mu_{n,-}^*$ barely
change. One hence can approximate the self energy by the value
$\mu_{n,+}\approx \mu_{\tilde n,+}\equiv\mu$ at the conical point.
Furthermore, the integration limits can be extended to $\pm
\infty$. The resulting integral
\begin{equation}
G=\frac{2e^2}{h}\int_{-\infty}^\infty\frac{2({\rm
Im}\,\mu)^2}{|\mu|^2 \cosh(2\delta\pi{\cal L}/{\cal W})-({\rm
Re}\,\mu)^2 +({\rm Im}\,\mu)^2}\,d\delta
\end{equation}
can be calculated exactly, and results in Eq.\
(\ref{eq:ggeneral}).

For the Fano factor (\ref{eq:fdef}) one is led to calculate a
similar integral $I\equiv\int|t_{nn}|^4\,dn$, which can be
evaluated using exactly the same steps as for the conductance.
The integral is then approximated as
\begin{equation}
I=\int_{-\infty}^\infty\frac{4({\rm Im}\,\mu)^4}{[|\mu|^2
\cosh(2\delta\pi{\cal L}/{\cal W})-({\rm Re}\,\mu)^2 +({\rm
Im}\,\mu)^2]^2}\,d\delta,
\end{equation}
which again can be evaluated exactly,
\begin{equation}
I=\frac{{\cal W}}{\pi{\cal L}}\frac{({\rm Im}\,\mu)^2}{({\rm
Re}\,\mu)^2}-\frac{{\cal W}}{\pi{\cal L}}\frac{{\rm Im}\,\mu[
({\rm Im}\,\mu)^2-({\rm Re}\,\mu)^2]}{({\rm
Re}\,\mu)^3}\arcsin\frac{{\rm Re}\,\mu}{|\mu|}.
\end{equation}
The final result for the Fano factor is given in Eq.\
(\ref{eq:fgeneral}).

% HS12: update references (and eliminate et al's)

\end{document}